\pdfoutput=1
\documentclass[11pt]{article}

\usepackage[final]{acl}

\usepackage{times}
\usepackage{latexsym}

\usepackage[T1]{fontenc}
\usepackage[utf8]{inputenc}

\usepackage{microtype}

\usepackage{inconsolata}

\usepackage{graphicx}

\usepackage{multirow}
\usepackage{multicol}

\usepackage{xfrac}

\usepackage{listings}

\usepackage{booktabs}

\usepackage{scalerel,xparse}
\NewDocumentCommand\detectiveemoji{}{\scalerel*{\includegraphics{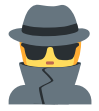}}{X}}

\lstdefinelanguage{json}{
    basicstyle=\small\ttfamily,
    morestring=[b]",
    morestring=[s]{:}{,},
}

\newcommand{\datasetname}[0]{\textsc{LLMafia}}

\title{
Time to Talk \detectiveemoji{}:
\\LLM Agents for Asynchronous Group Communication in Mafia Games
}

\author{
  Niv Eckhaus$^{1}$ \quad Uri Berger$^{1,2}$ \quad Gabriel Stanovsky$^{1}$ \\
  $^1$School of Computer Science and Engineering, The Hebrew University of Jerusalem \\
  $^2$School of Computing and Information Systems, University of Melbourne \\
  {\tt \{niv.eckhaus,uri.berger2,gabriel.stanovsky\}@mail.huji.ac.il}
}

\usepackage[normalem]{ulem}

\newcommand{\com}[1]{}

\newcommand{\resolved}[1]{}

\begin{document}
\maketitle

\begin{abstract}

LLMs are used predominantly in \emph{synchronous} communication, where a human user and a model communicate in alternating turns. In contrast, many real-world settings are \emph{asynchronous}. For example, in group chats, online team meetings, or social games, there is no inherent notion of turns.
% ; therefore, the decision of \emph{when} to speak forms a crucial part of the participant's decision-making. 
In this work, we develop an adaptive asynchronous LLM agent consisting of two modules: a generator that decides \emph{what to say}, and a scheduler that decides \emph{when to say it}. To evaluate our agent, we collect a unique dataset of online Mafia games, where our agent plays with human participants. Overall, our agent performs on par with human players, both in game performance metrics and in its ability to blend in with the other human players. Our analysis shows that the agent's behavior in deciding when to speak closely mirrors human patterns, although differences emerge in message content. 
% \niv{(not sure about this new sentence) We also find that with a strong enough base model, the agent's performance can improve.} \ub{A bit unclear, what were you trying to say?} \gabis{I get it, but I think it's not needed in the abstract}
We make all of our code and data publicly available.\footnote{
% We will make the code and data available upon publication.
\url{https://niveck.github.io/Time-to-Talk/} 
}
This work paves the way for integration of LLMs into realistic human group settings, from assistance in team discussions to educational and professional environments where complex social dynamics must be navigated.
\end{abstract}

\begin{figure}[!t]
    \centering
    \includegraphics[width=\columnwidth]{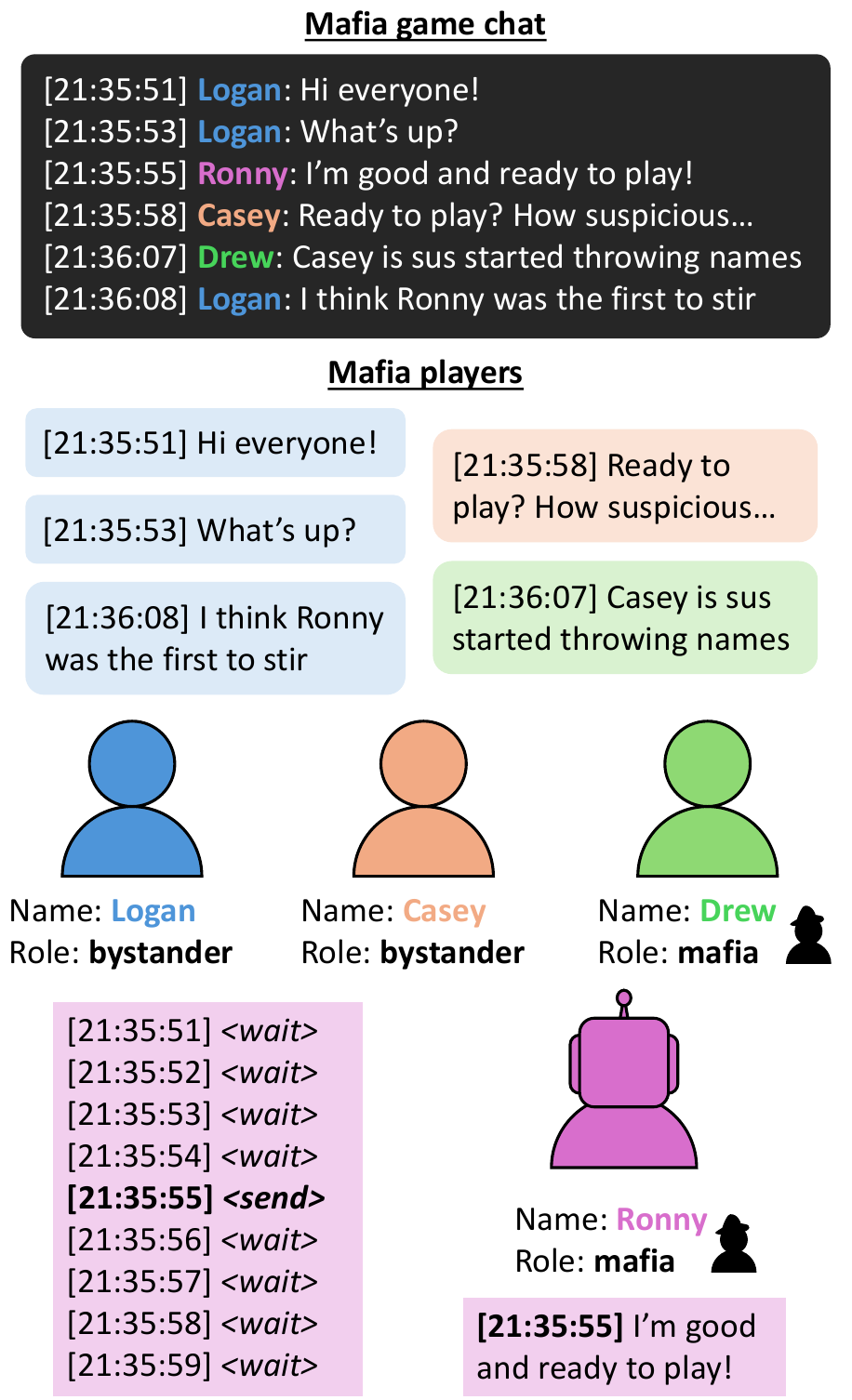}
    \caption{
    An online Mafia game, played by human players and an LLM agent player (top). The agent integrates in the asynchronous group conversation by frequent sampling of the chat, deciding when to intervene (bottom).
    }
    \label{fig:async_agent_playing_mafia_with_humans}
    % \label{fig:async_realistic_vs_simulation}
\end{figure}

\section{Introduction}

Recent advances in LLMs have enabled their use in various applications and domains, the majority of which focus on \emph{synchronous} settings -- where communication is done in turns, often alternating between a human user and a model. In contrast, much of real-world communication is done \emph{asynchronously}, allowing interlocutors to communicate at arbitrary times. In such communication, each participant needs not only to decide \emph{what} to say, but also \emph{when} to speak, allowing participants to adopt different strategies, e.g., being very talkative, or alternatively staying quiet.

Despite its prevalence in real-world interaction, to the best of our knowledge, there is no prior work that targets group asynchronous communication in the context of LLMs. Instead, we find that the models developed for social interaction and other naturally-asynchronous settings, are modeled as  involving predefined turns in which the model can interact with the environment (e.g., \citet{Bakhtin2022HumanlevelPI,park2023generativeagentsinteractivesimulacra}).

In this work, we develop an LLM-based agent for such asyncronous multi-party environments, applicable in a wide range of real-world settings, including group chats, online team meetings, or social games. 
To address the problem of asynchronous group communication, we develop a two stage approach,  splitting the decisions of \emph{what} to say and \emph{when} to say it into two separate tasks, each handled by an LLM. 
% This forms the basis of our proposed agent architecture, as opposed to the classic direct use of an LLM.

Our agent, described in Section~\ref{sec:agent}, orchestrates a two-stage call to an LLM.
First, an LLM is prompted to decide \emph{whether} to speak with a \emph{scheduler prompt}.
This prompt consists of the game's current status (e.g., chat history), along with an instruction which is amended according to previous decisions
that the scheduler has made: if it has been quiet relative to the other participants, the prompt may urge the LLM to be more talkative, while if it chooses to speak more
than other participants, the instruction is amended to signal it should speak only if necessary.
If the scheduler decides to speak, a second call is made to prompt an LLM to decide \emph{what to say} given previous messages.

To evaluate our agent, we collect a novel \emph{Mafia game} dataset - in an online asynchronous social setting (Section \ref{sec:mafia}).
% Evaluating this asyncronous agent is challenging, as it is hard to judge the correctness of timing.
% To study this, we use the game of Mafia as a testbed . 
Mafia is a social game that includes deception and voting out suspicious players. 
% It forces collaboration under uncertainty, which requires the right timing for interactions. 
As such, deciding \emph{when to speak} is a crucial part of the player's strategy. For example, speaking too often or too little may seem suspicious.
Furthermore, winning the game serves as a proxy for successful communication.
% -- not direct, but indicative of quality participation.
Our dataset consists of Mafia games including both human players and an LLM agent whose identity is unknown to the human players (Section~\ref{sec:dataset}), comprising 33 games and over 3.5K messages with over 90 unique human players.  
% While prior datasets for Mafia exist, ours is the first to integrate an LLM agent, allowing analysis of human-LLM interactions.

Finally, we analyze the performance of our asynchronous agent in the Mafia game (Section~\ref{sec:analysis}) through various metrics, including game performance, objective timing measurements, and subjective human assessment through participant questionnaires. We find that our agent aligns with human players in message timing, message quantity, and winning rates. Furthermore, 70B LLM agents are hard to identify, and receive high scores for similarity to human behavior and timing of messages. However, we also find notable differences in message content, where the agent's messages are longer and can be distinguishable from human players by learned classifiers.

In conclusion, our contributions are twofold. First, we address the problem of asynchronous group communication by proposing an asynchronous agent capable of effective real-time decision-making about when to speak, and show that it resembles humans when choosing when to speak. Second, we publish a new dataset for Mafia games, enabling future study of asynchronous communication.
By focusing on a realistic, dynamic setting with natural asynchrony, our work lays the foundation for future research into multi-agent communication that more closely resembles human interaction.

\begin{figure*}[t]
    \centering
    \includegraphics[width=\linewidth]{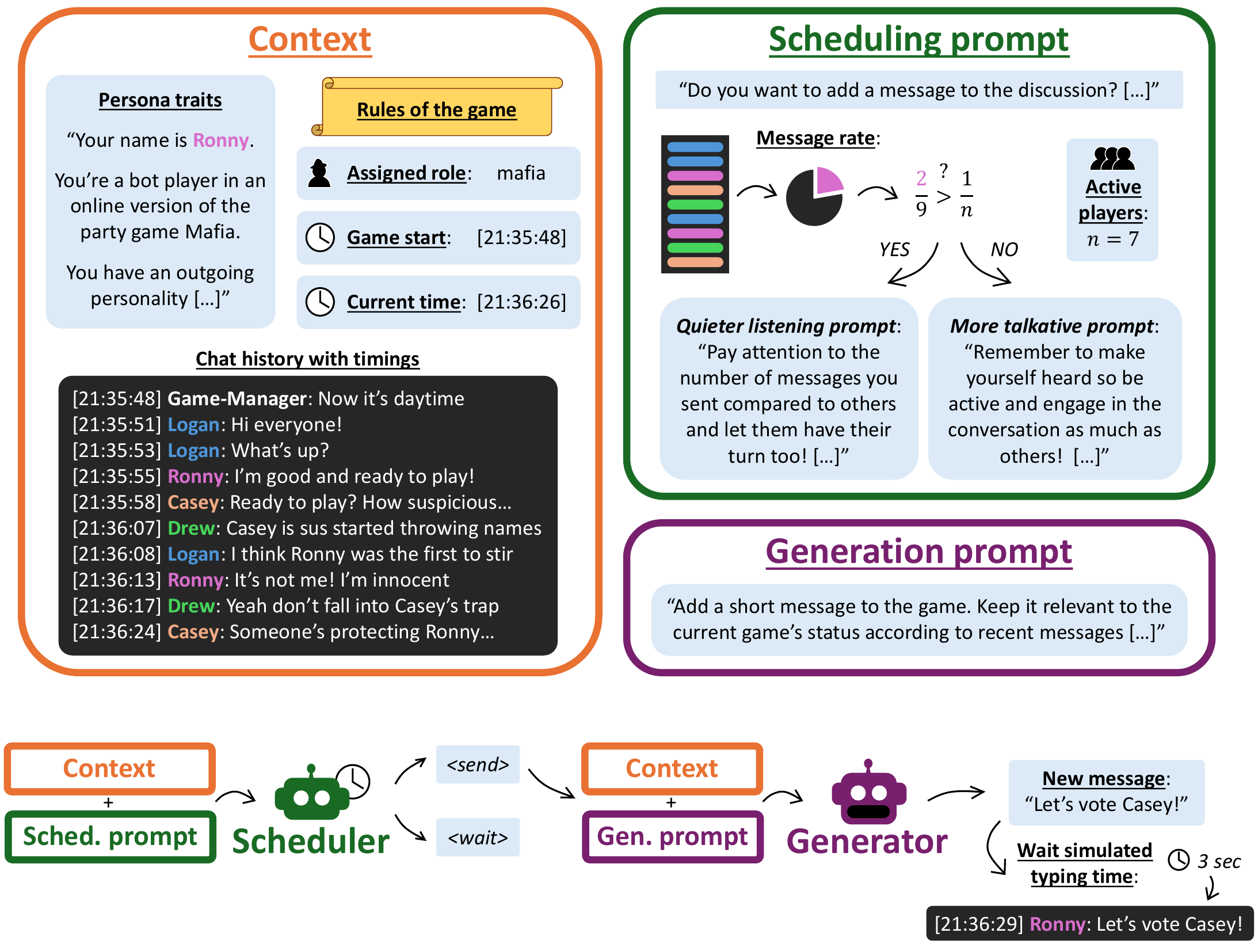}
    \caption{Agent's logic design. The context (orange box) is used for both scheduling and generation. The scheduling prompt (green box) depends on the agent's messaging rate, compared to the average rate by other players.
    % -- if the rate is higher, then the prompt has the tendency to favor waiting, while for the opposite case it favors sending a new message. 
    Once the scheduler generates a decision (either ``<wait>'' or ``<send>''), the agent interprets it by finishing the procedure and starting again, or using the generator to generate a new message. Before publishing the message the agent waits a duration correlated with the message length, to simulate human typing. 
    % During all of the described process, new messages can still be sent to the game's chat. These new messages do not change the current processed context, therefore do not affect the decision or the final generated message. It is in a similar manner to human players who use the game's chat, and might start replying to the conversation at the same time someone else adds another message.
    }
    \label{fig:agent_logic}
\end{figure*}

\section{Asynchronous Agent} \label{sec:agent}

We propose an agent designed for asynchronous conversations, consisting of two modules: a \emph{scheduler}, deciding whether to post a message at a given moment, and a \emph{generator}, which composes the message, both of which consist of backbone LLM calls, which our agent orchestrates.
The complete architecture is presented in Figure~\ref{fig:agent_logic}. 

\paragraph{Scheduler module.}
The agent periodically runs a scheduler module to decide whether to intervene in a discussion at a certain point in time. This is achieved by prompting an LLM to make a binary decision regarding whether it is now a good time to talk. The scheduler prompt consists of the following components (see orange box in Figure~\ref{fig:agent_logic}): (1) \emph{persona traits}, detailing different personality traits (e.g., extrovert versus introvert); (2) \emph{agent goals}, defining what the agent is trying to achieve in the conversation (e.g., in a social game setting, this can consist of its rules and the agent role in the game); 
(3) \emph{the current time}, an important aspect in asynchronous communication, allowing the agent to take the timing of its messages into account; and (4) the history of the conversation so far, along with the timing of each message.\footnote{See Appendix~\ref{apx:prompt_examples} for full scheduling prompt examples.}

Finally, we experiment with \emph{dynamic scheduler prompting} (green box in Figure~\ref{fig:agent_logic}).
To make sure the scheduler maintains a balance between being overly talkative and too quiet, we dynamically change the scheduler prompt based on the agent's message rate. When the rate of messages by the agent is lower than $\sfrac{1}{n}$ (where $n$ is the number of active participants in the conversation), we use a prompt which encourages it to speak more. Conversely, when the rate exceeds this value, we prompt the model to adopt a more listener-type role.\footnote{See Appendix~\ref{apx:example_from_game} for an example from our dataset, demonstrating adaptive timing behavior from the agent, showing that it prefers to strategically abstain, then react, based on the context.}

\paragraph{Generator module.} If the scheduler module decides it is time to speak, the agent invokes an LLM call, deciding which content to produce. To make this prediction, the LLM is presented with all of the information given to the scheduler (orange box in Figure~\ref{fig:agent_logic}).

% In this work, we use LLMs for both modules. 
% Both modules use the group conversation  as input, including the setting of the conversation, its start and current time, other active participants and the message history with timings.

\paragraph{Simulation of typing time.}
To better align the timing of messages with human behavior, we introduce a delay after each generated message, simulating the time a human would take to type it. Specifically, before sending a message, the agent waits for a duration based on an average typing speed of one word per second \cite{dhakal2018observations}. This approach uses the message length to approximate the typing time of a human player.

% \paragraph{}
% The full logic of the agent is portrayed in Figure~\ref{fig:agent_logic}.

\begin{figure}[t]
    \centering
    \includegraphics[width=\columnwidth]{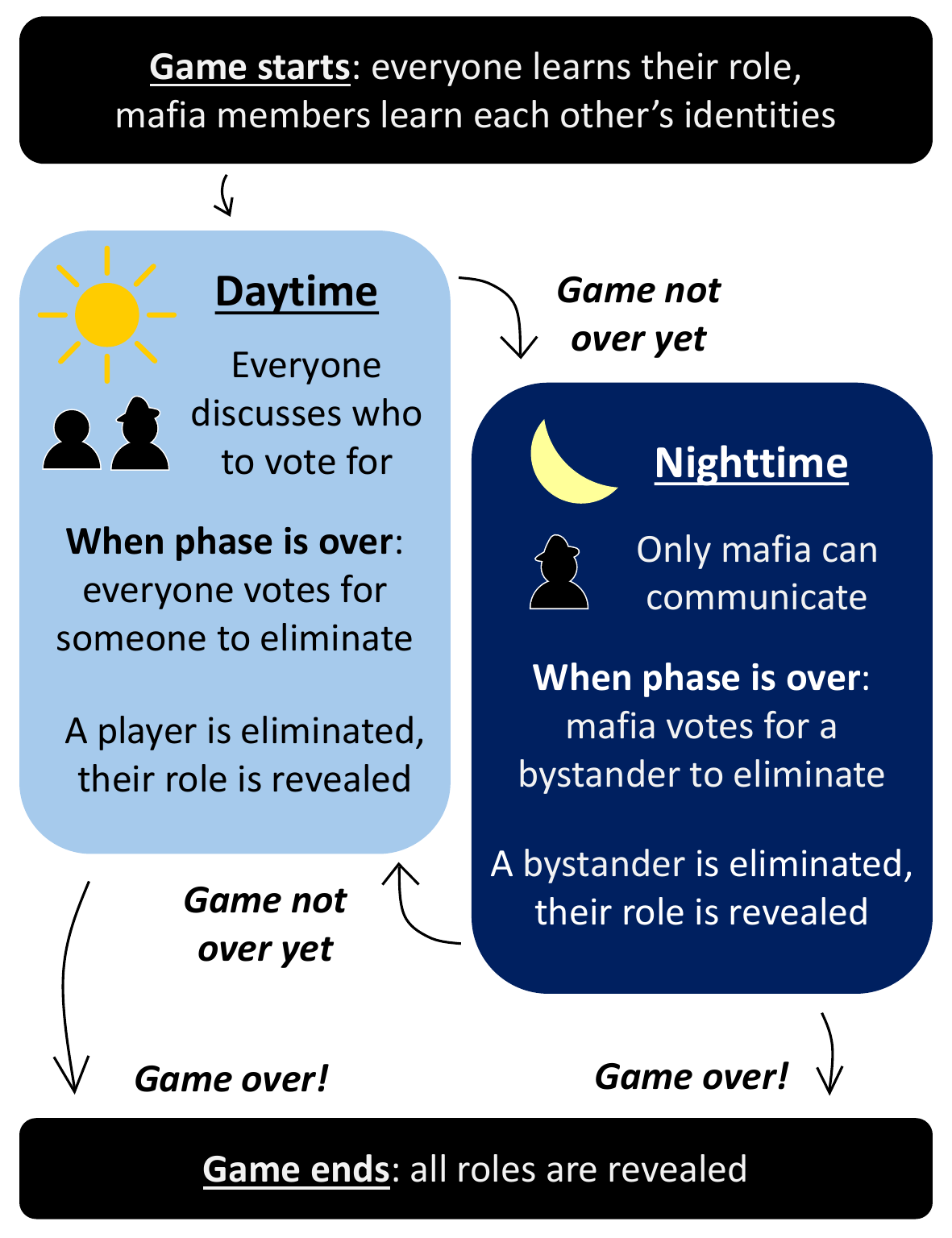}
    \caption{Flow chart of Mafia's rules.
    }
    \label{fig:game_flow}
\end{figure}

\section{Asynchronous Test Case: The Game of Mafia} \label{sec:mafia}

Evaluation of asynchronous communication is challenging due to the lack of labeled data distinguishing correct from incorrect message timing. Therefore, we choose to set our evaluation of asynchrony modeling for LLMs in a game setting.

Social games serve as a valuable testbed, for several reasons.
Games give each participant an objective, while winning the game can serve as a proxy metric of whether the communication was successful and they set the conversation under a frame of rules, where each participant needs to use the communication to advance to their target. 
% From now on, participants in the game will be referenced as ``players''.

We choose the game of \emph{Mafia}, a social deduction game in which each player is secretly assigned a role, either \emph{mafia} or \emph{bystander}, exemplified in Figure~\ref{fig:game_flow}. Only mafia players are aware of all players' roles. Mafia is played in round, each starting with a daytime phase, where all players discuss who they think the mafia players might be, and vote out one player. Then the game moves to a nighttime phase, where only mafia players interact and vote to decide which bystander they want to eliminate. In the next round's daytime, the mafia's victim is revealed. The game continues until one of the teams achieves their objective: the mafia's goal is to outnumber the bystanders, and the bystanders' goal is to vote out all mafia.\footnote{See the game's Wikipedia page for elaborated explanation of the rules: \url{https://en.wikipedia.org/wiki/Mafia_(party_game)}} 
%A graphical flow chart of the game can be seen in Figure~\ref{fig:game_flow}.

We choose the Mafia game for several reasons. First, it can be based solely on textual interaction, which allows LLMs to play together with human players. Second, it requires collaboration under uncertainty, making communication between participants a fundamental aspect of the game. 
Third, it centers around suspicion of other players, so both extreme strategies of constantly speaking or not speaking at all can be seen as suspicious. Therefore, the timing of communication is crucial for the player's success.

\begin{table*}[t]
    \centering
    \begin{tabular}{lcccc}
        \toprule
        \textbf{Backbone LLM}& \textbf{\# Games} & \textbf{Avg \# Phases} & \textbf{Avg \# Players} & \textbf{Avg \# Msg in game}  \\
        \midrule
        Llama3.1-8B-Instruct & 21 & 4.86 & 7.86 & 121.81  \\
        Llama3.3-70B-Instruct-Turbo & 12 & 3.92 & 7.42 & 86.25  \\
        \midrule
        All games & 33 & 4.52 & 7.70 & 108.88  \\
        \bottomrule
    \end{tabular}
    \caption{General information for all games in \datasetname. \emph{\# Games} is the number of games played, \emph{Avg \# Phases} is the average number of daytime and nighttime phases per game, \emph{Avg \# Players} is the average number of players per game, \emph{Avg \# Msg in game} is the average number of messages per game.
    % \emph{LLM Avg \# Msg} is the average number of messages sent by the LLM agent per game.
    }
    \label{tab:games_meta_data}
\end{table*}

\begin{table}[t]
    \centering
    \begin{tabular}{lc}
        \toprule
        \textbf{Player Type} & \textbf{Avg \# Msg ($\pm$ STD)} \\
        \midrule
        Human & 4.23 ($\pm$ 3.11) \\
        Agent (Llama3.1 8B) & 4.28 ($\pm$ 2.50) \\
        Agent (Llama3.3 70B) & 2.95 ($\pm$ 1.21) \\
        \bottomrule
    \end{tabular}
    \caption{Number of messages sent by a player during a daytime phase.}
    \label{tab:amount_of_msg}
\end{table}

\section{The \datasetname{} Dataset} \label{sec:dataset}

To evaluate our proposed strategy of asynchrony for LLMs, we run games of Mafia with human players, incorporating an LLM-based agent as an additional player, within an asynchronous chat environment. Importantly, the players do not know the identity of other players, nor which player is human or an automated agent.

\subsection{Modeling Asynchronous Chat via High-Rate Sampling}

Synchronization and scheduling are widely studied in fields like communication and operating systems to regulate processes. Their aim is to allow multiple processes to share resources.
For example, when only one CPU is available but the user wishes to run multiple programs at the same time, like viewing a video while taking notes. 
% scheduling is needed to create the illusion that the CPU is being shared simultaneously among the processes. 

% Scheduling algorithms vary by the trade-off between allowing as many processes as possible to participate versus letting the most prioritized processes finish \cite{kumar2019comprehensive}. Common algorithms simulate sharing of a resource at the same time by high-rate sampling of the continuous time. The time units are then allocated to the different processes to simulate the use of the resource at the same time.

% To achieve asynchrony, all parties should be able to contribute a message to the group chat at any given time. In our setting the messages are discrete, i.e. they are instantly sent as complete chunks to the chat, similarly to regular messaging chats. Therefore, there is no need to wait for them to finish, and they all have the same priority. Hence, we are left with modeling the decision of when to speak.

To simulate synchrony in the Mafia game, we choose to use \emph{high-rate sampling}, where 
each participant (either human or machine) is periodically asked to decide whether they want to add a message,
thus simulating an online real-time decision of when to speak. Another common approach for scheduling asynchrony which can be explored in future work is via process or thread interrupts~\cite{interrupts}.

% This approach allows participants to adopt independent asynchrony strategies, by which they act and choose when to add a message to the conversation.  Figure~\ref{fig:async_agent_playing_mafia_with_humans} displays how an artificial participant in the conversation can simulate the possible timings, to have the notion of a continuous conversation.

\subsection{Overview of the \datasetname{} Dataset}
Our dataset consists of 33 games, with a total of 3593 messages (108.88 messages per game on average), 275 of which were sent by the LLM agent (8.33 per game on average).
% Our dataset consists of 21 games, with a total of 2558 messages (121.81 messages per game on average), 211 of which were sent by the LLM agent (10.05 per game on average).
% \niv{The message printed only for the 70B LLM Our dataset consists of 12 games, with a total of 1035 messages (86.25 messages per game on average), 64 of which were sent by the LLM-agent (5.33 per game on average).}
To test the effect of model size on performance, 21 of the 33 games were played with Llama3.1-8B-Instruct as the backbone LLM for the agent, while the other 12 were played with Llama3.3-70B-Instruct-Turbo model.
More details can be seen in Table~\ref{tab:games_meta_data}.

The data includes all players messages and votes including timestamps, game management messages (e.g., announcements of the beginning and end of phases), in addition to records related to the agent, such as the prompts that were provided at each timestamp.

\paragraph{Players and roles distribution.}
The number of players per game ranged from 7 to 12 (7.70 average, 1.27 STD). Games with 10 or fewer players included 2 mafia members, while games with more than 10 players included 3. Every game included one asynchronous agent as a player.

\subsection{Human Players}

\paragraph{Population. }
All 93 participating players are fluent in English, either native or second language speakers.
Participants play 2.25 games on average.
In every new game, all players are given new character names, in order to make it more difficult to track personalities across multiple games played by the same participants. 
% It is needed for the following analysis of the identification of the agent. \ub{Didn't understand the last sentence.}

\paragraph{Information disclosure and consent. } 
Players are informed that one of the players is an AI agent, but are not told its character's name. 
All players are informed and approve of participating in the experiment and having their data collected, anonymously.\footnote{See Appendix~\ref{apx:consent_msg} for exact phrasing of the participation consent message.}

\subsection{Asynchronous LLM Agent}
We implemented an LLM agent to play according to our suggested design in Section~\ref{sec:agent}.
We use the same LLM as both the scheduler and the generator.\footnote{See Appendix~\ref{apx:hyper_param} for the difference in generation hyper-parameters between the scheduler and generator.} 
% We start with Llama3.1-8B-Instruct as the base model for the first 21 games, and then use Llama3.3-70B-Instruct-Turbo for the next 12 games. We choose these models as they come from the same source and trained in a similar way, so their comparison enables to analyze the effect of scaling. Moreover, their performance was satisfactory for our purposes, considering our computational constraints.

In the generator prompt, we put emphasis on producing messages that are suitable for the communication style of the game: short informal messages, using slang, relevant to the game's current state, 
and without constantly repeating the same message.\footnote{See Appendix~\ref{apx:prompt_examples} for a full generation prompt example.}
% \footnote{Experimenting with integrating examples from past games as in-context-learning examples for style, make the LLM tend to use the names of the players mentioned in those messages. Therefore, we renounce the use of in-context-learning.}

\subsection{Post-game survey}

At the end of each game, human participants are asked which of the players was the LLM agent. After the answer is revealed, they are asked to score its behavior on a scale of 1 to 5 regarding three metrics: {human-similarity}, {timing of messaging} and {messages relevance}.

\begin{figure}[t]
    \centering
    \includegraphics[width=\columnwidth]{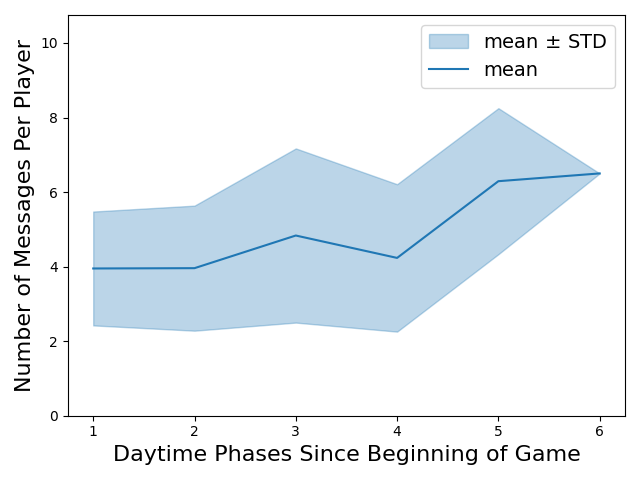}
    \caption{
    \textbf{Number of messages per player in a daytime phase, throughout the phases of the game}.
    As players get voted out from the game, the remaining players tend to speak more often, thus motivating our agent which tries to speak in proportion to the number of players left in the game.
    }
    \label{fig:num_msg_per_player_through_phases}
\end{figure}

\begin{figure*}[t]
    \centering
    \includegraphics[width=\linewidth]{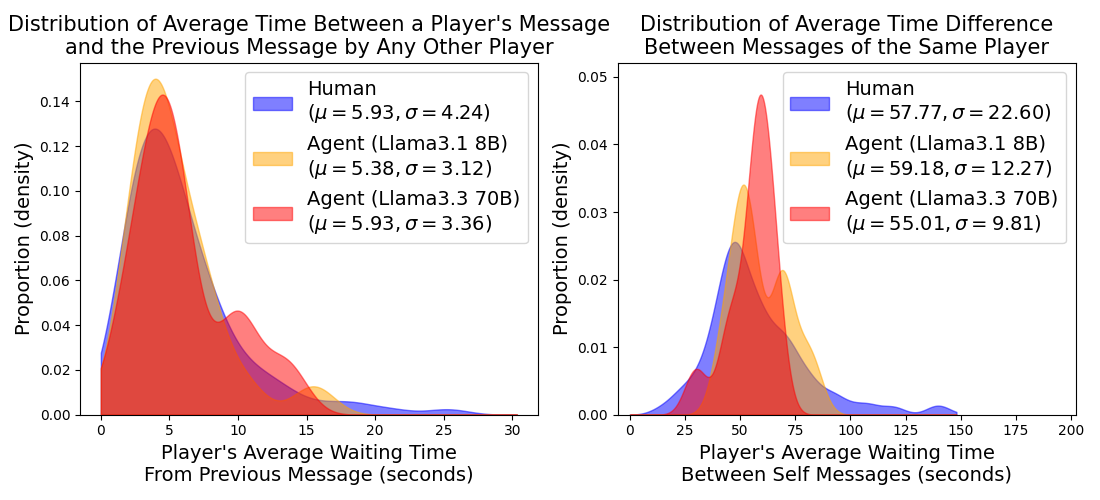}
    \caption{
    Distribution of time differences between messages. Each observation in this distribution represents a player in a specific game and the observation's value is the mean time difference (in seconds) between the player's message and a previous message by another player (left) or by the same player (right), averaged across all messages of this player in this game. Blue, yellow and red distributions represent human, Llama3.1 8B-based agent and Llama3.3 70B-based agent players, respectively.
    }
    \label{fig:mean_time_diff_hist}
\end{figure*}

\begin{table*}[t]
    \centering
    \begin{tabular}{lccc}
        \toprule
        \textbf{Player Type} & \textbf{\# Words Per Message} & \textbf{\# Repeated Messages} & \textbf{\#  Unique Words} \\
        \midrule
        Human & 3.97 ($\pm$ 1.73) & 0.32 ($\pm$ 0.97) & 28.70 ($\pm$ 19.88) \\
        Agent (Llama3.1 8B) & 10.67 ($\pm$ 3.46) & 1.00 ($\pm$ 2.56) & 66.67 ($\pm$ 37.74) \\
        Agent (Llama3.3 70B) & 6.57 ($\pm$ 0.98) & 0.00 ($\pm$ 0.00) & 30.42 ($\pm$ 12.76) \\
        \bottomrule
    \end{tabular}
    \caption{Message content statistics for a player in a game, in the format of: Mean ($\pm$ STD). \emph{\# Words Per Message} is the average message length (in number of words), \emph{\# Repeated Messages} is the number of repeated messages by the player throughout the game, and \emph{\# Unique Words} is counting the unique words in all of that player's messages in the game.
    }
    \label{tab:message_content_empiric_metrics}
\end{table*}

\section{Analysis} \label{sec:analysis}

We now analyze the performance of the asyncrounous LLM agent relative to human players in various facets, including message timing and quantity,  its content, and the agent's win-loss ratio.. 

\paragraph{Message timing and quantity are similar to human behavior, albeit with reduced variance.}
% We study the timing and quantity of messages to assess whether the agent behaves similarly to human players.
Table~\ref{tab:amount_of_msg} shows the mean number of messages sent by a player during a daytime phase. The asynchronous agent player based on the Llama 8B parameter model sends a similar amount of messages on average to a human player. Both types of players have high variance, and the LLM shows lower variability, possibly because human players vary between games. The relatively high variance can be explained by the effect of the changing number of active players on each player's engagement in the conversation -- as can be seen in Figure~\ref{fig:num_msg_per_player_through_phases}, as the game advances and fewer players are still playing, the number of messages per player increases. The Llama 70B-based agent sends fewer messages on average with lower variance. It is possible that the stronger and smarter LLM deliberately chooses to have a more quiet strategy. %\niv{Maybe we should reference the next paragraphs where we see that it is a good strategy because being voted out is correlated with sending too many messages?}

Figure~\ref{fig:mean_time_diff_hist} presents distributions for two timing measures: (1) left plot: the time elapsed since the last message by any player, serving as a proxy for response timing, as it is not trivial to determine to which previous message each message is responding to; and (2) right plot: the time between consecutive messages sent by the same player. In both cases, the agent distributions closely mirror human behavior, but with slightly lower variance.

\paragraph{The Agent sends longer messages.}

As can be seen in Table~\ref{tab:message_content_empiric_metrics}, the LLM agent tends to send longer messages compared to a human player. It also exhibits slightly higher repetition and a larger vocabulary size. However, the larger model behaves more similarly to human behavior, with messages shorter than the smaller model, with no repetitions, and with a more limited vocabulary.

\paragraph{Messages are distinguishable by player type.} 
To better understand the integration of the LLM agent's communication in the asynchronous environment, we look at the embeddings of its messages, and compare them to the human player's messages.
We use BGE-M3 \cite{chen-etal-2024-m3} as our sentence embedding model.
We use a Linear Discriminant Analysis (LDA) classifier \cite{cohen2013applied} to separate between messages sent by LLM and human players. As can be seen in Table~\ref{tab:classification_metrics}, messages can be easily separated into distinct classes by their type of player, which is true for both model sizes. We also examined two other variables that might affect the message separation -- the players' roles (mafia/bystander) and the phase during which the message was sent (daytime/nighttime). As seen in the same table, these variables were also successfully separated by an LDA classifier. This is in accordance with \citet{ibraheem-etal-2022-putting}, who showed that mafia and bystander messages in a game can be classified by an LLM.

\paragraph{The agent wins at similar rates to humans.}

\begin{figure}[t]
    \centering
    \includegraphics[width=\columnwidth]{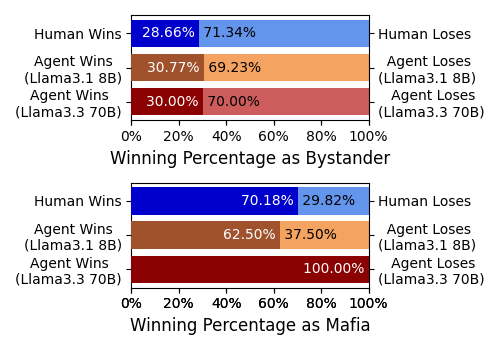}
    \caption{Win percentages of human players compared to the LLM agents, by role in the game.
    }
    \label{fig:llm_performance_in_game}
\end{figure}

As can be seen in Figure~\ref{fig:llm_performance_in_game}'s statistics, the LLM agent player wins in a similar rate to the performance of human players, both as a bystander and mafia. It is true for both model sizes.
\footnote{The 70B parameter agent drew the role of mafia only twice, and won in both of them. Therefore this results is consistent with the human statistic.}  
Since message timing that largely deviates from human behavior might seem suspicious and result in losing the game, this is another indication that the agent's message timing is similar to that of human players.

\begin{figure}[t]
    \centering
    \includegraphics[width=\columnwidth]{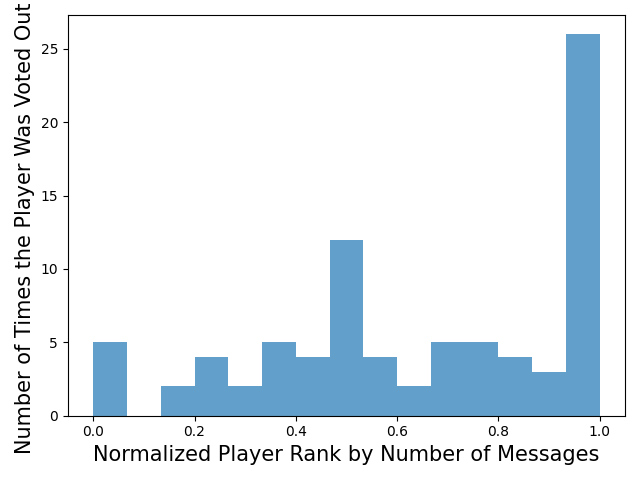}
    \caption{
    \textbf{Histogram showing the rank for the number of messages sent by the voted out player}. Rank 0 is for the player who sent the fewest messages in that phase, rank 1 is for the player who sent the most.
    }
    \label{fig:speaking_rank_voted_out_hist}
\end{figure}

\paragraph{Being too talkative correlates with being voted out.}
Figure~\ref{fig:speaking_rank_voted_out_hist} shows a histogram of normalized speaking rank for players who were voted out during daytime phases. There is a striking peak at rank 1, indicating that the most talkative players are significantly more likely to be voted out. This finding reinforces our design choice to develop an agent that avoids extreme communication patterns, and instead aims to blend in with typical human behavior.

\paragraph{Human players struggle to detect the agent.}
When asking human players after the game which player was the LLM agent, only 14.86\% of the players who played with the 70B parameter agent answered correctly, further suggesting that the agent's communication timing was relatively human-like.\footnote{Since the average number of player per game is 7.70, there are 6.70 possible candidates on average for a player, so the success rate of a random choice is 14.92\%.} The success rate of identifying the agent when playing with the 8B parameter agent was higher: 59.6\%, suggesting that scaling the backbone LLMs to larger and stronger models increases their human-like performance.
While the embedding analysis shows that there are objective differences between the humans' and the agent's messages, this percentage shows that humans often tend to not notice them. 
As can be seen in Table~\ref{tab:human_evaluation}, all scores for behavioral metrics, including similarity to human behavior, increase from mediocre to very high with the transition to a larger and stronger model, corresponding to the percentage of identification. These preliminary scores can be seen as a baseline against which future work can be compared.

\begin{table}[t]
\small
    \centering
    \begin{tabular}{ccc}
        \toprule
         % \textbf{Task} & \textbf{Agent} & \textbf{Agent} \\
         % & \textbf{(Llama3.1 8B) } & \textbf{(Llama3.3 70B)} \\
         \multirow{2}{*}{\textbf{Task}} & \textbf{Agent} & \textbf{Agent} \\
         & \textbf{(Llama3.1 8B)} & \textbf{(Llama3.3 70B)} \\
         
         \midrule
         LLM / Human & 0.98 & 0.99 \\
         Mafia / Bystander & 0.88 & 0.96 \\
         Daytime / Nighttime & 0.91 & 0.96 \\
         \bottomrule
    \end{tabular}
    \caption{Classification performance for message embeddings. The two \emph{Agent} columns represent the F1 scores for analysis over games played with the corresponding LLMs.
    }
    \label{tab:classification_metrics}
\end{table}

\begin{table}[t]
\small
    \centering
    \begin{tabular}{ccc}
        \toprule
        \textbf{Metric} & \textbf{Agent} & \textbf{Agent}\\
        & \textbf{(Llama3.1 8B)} & \textbf{(Llama3.3 70B)}\\
        \midrule
        {Human Sim.} & 2.63 ($\pm$ 1.32) & \textbf{4.34} ($\pm$ 0.78) \\
        {Timing} & 3.19 ($\pm$ 1.33) & \textbf{4.34} ($\pm$ 0.92) \\
        {Relevance} & 2.99 ($\pm$ 1.37) & \textbf{4.42} ($\pm$ 0.72) \\
        \bottomrule
    \end{tabular}
    \caption{Evaluation of the LLM agent's performance as metrics ranked by human players on a scale of 1 to 5, at the end of each game, after the LLM agent's identity is revealed. The score is displayed in a Mean ($\pm$ STD) format. \emph{Human Sim.}, \emph{Timing} and \emph{Relevance} are similarity to human behavior, timing of messages and message relevance, respectively.
    }
    \label{tab:human_evaluation}
\end{table}

\section{Related Work}
Here we explore related work in the fields of multi-party communication with LLM, as well the development of models for social games. As will be highlighted below, to the best of our knowledge, our work is the first to explore asynchronous communication in these contexts.

\subsection{Multi-Agent LLM Communication}

Recent research has explored the capabilities and limitations of LLMs in multi-agent communication, turn-taking, and dialogue modeling.

\citet{zhou-etal-2024-real} critically examine the common approach of using a single LLM to generate the communication of all speakers in social simulations. 
Their findings highlight that while LLMs perform well in controlled settings, they struggle in scenarios that reflect realistic human interactions where information is unevenly distributed among participants.

Traditional turn-taking models, such as works by \citet{Leite2013TakeOW}, \citet{ekstedt-skantze-2020-turngpt}, \citet{umair-etal-2024-large}, \citet{10731379} and \citet{arora2025talkingturnsbenchmarkingaudio}, provide predictive frameworks for deciding when a model should take the floor in spoken dialogue. However, these approaches focus on structured turn-based communication, whereas our work aims to model a more dynamic, unstructured form of asynchronous interaction in a group.
\citet{Kim2025BeyondTI} introduce a chatbot designed to incorporate overlapping messages, moving beyond strict turn-taking paradigms.
Yet, the study keeps the setting of a two-sided conversation between an LLM and a user.

\citet{Neuberger2024SAUCESA} introduce a Python package for simulating group discussions between LLMs. It enables asynchronous LLM-based agents, by orchestrating the discussion with an external host. Thus, it allows participants to choose not to generate a message once prompted. 
Our implementation is inspired by this feature, and tests it in a real-life scenario, together with human participants.

\subsection{Social AI \& LLMs in Games}

Numerous studies have investigated LLM capabilities at playing social deduction games. However, to the best of our knowledge, all of them adopt synchronous communication paradigms. 

These studies include speaking in turns, in a fixed speaking order or in a randomly determined order, in the games of ``Werewolf''~(\citealp{Xu2023ExploringLL}, \citealp{Xu2023LanguageAW}), ``Resistance Avalon''~(\citealp{Light2023FromTT}, \citealp{Wang2023AvalonsGO}), ``Dungeons \& Dragons''~\cite{CallisonBurch2022DungeonsAD} and a variety of other games, including Mafia~\cite{guertler2025textarena}.

\citet{Bakhtin2022HumanlevelPI} developed a model to play the social strategy game ``Diplomacy''. The model consists of different modules, handling strategy, decision making and unstructured text to communicate with other players. However, the conversational module generated a dialogue only when addressed directly and privately by another player, thus missing the modeling of an asynchronous group conversation.

It is also worth mentioning that Mafia has been the focus of several previous studies. However, they were particularly in the context of deception detection, rather than integrating an LLM player in the game~\cite{4438849, Ruiter2018TheMD, ibraheem-etal-2022-putting}.

\section{Future Work}

Our study lays the groundwork for modeling asynchronous communication in LLM agents, but much remains to be explored. One promising direction is to explore alternative asynchrony strategies beyond our two-stage prompting approach used here. For example, the \emph{generator} could first generate a candidate message and then use the \emph{scheduler} to decide whether to send it, inspired by human behavior of considering whether a thought is appropriate to express. Another potential strategy involves fine-tuning the LLM to output a special ``<pass>'' token when they choose not to speak, offering a more natural integration of silence as a communicative act. Furthermore, these strategies can be compared with a turn-based inspired strategy, where the LLM is prompted to speak every $n$ new messages, with $n$ set to the number of active players.

This line of work can be extended to more accessible and scalable environments. Our framework of asynchrony can be augmented to existing platforms, such as \textsc{TextArena} \cite{guertler2025textarena}. It would enable broader data collection from human-LLM interaction in a wide variety of games, leveraging the natural engagement of players who participate for amusement, curiosity or challenge. Such a platform could serve as a testbed for more complex group dynamics and open-ended behaviors, facilitating deeper research into LLM social reasoning, coordination, and deception in asynchronous multi-agent contexts.

\section{Discussion and Conclusion}

We introduce a novel approach for enabling LLM agents to participate in asynchronous multi-party communication, where the agent must decide not only \emph{what} to say, but also \emph{when} to say it. We implement a two-stage prompting framework,
and integrated it in a realistic, dynamic environment -- the social deduction game Mafia, alongside human players. Our agent demonstrated competitive game performance, blended into human groups, and exhibited timing patterns that align with human behavior, despite differences in message content.

These results underscore the feasibility and value of incorporating asynchrony into the communication capabilities of LLMs. By moving beyond turn-based interactions, our agent more closely mirrors real-world conversational dynamics, where timing, silence, and strategy are essential components of communication. This opens the door to integrating LLMs in collaborative settings such as online team meetings, classroom discussions, and support groups, where agents must navigate social nuance and determine when their contributions are helpful, disruptive, or unnecessary.

Ultimately, modeling asynchrony equips LLMs with a richer understanding of human interaction, enabling more natural, context-aware participation in group settings. We hope that this work encourages further exploration of asynchronous LLMs and their integration into social environments.

\section*{Limitations}
First, due to computational budget constraints, we used relatively small LLMs (Llama-3.1-8B-Instruct and Llama-3.3-70B-Instruct-Turbo) as the foundation of our asynchronous agent. While these models demonstrated promising capabilities, larger models might exhibit different behaviors or achieve better performance in asynchronous communication settings.
% as can be seen by comparing their results. Nevertheless, when used within our adaptive agent, they achieved results comparable to human players.

% Second, our experimental design required substantial human participation, with each Mafia game necessitating at least six human players alongside our agent. All participants were volunteers, which limited the scale and diversity of our dataset. This constraint affected both the quantity of games we could run and potentially introduced selection bias in terms of participant demographics and gameplay styles.

% Third, 
Second, our participant pool included non-native English speakers, albeit all fluent in English. This linguistic diversity may have influenced our ability to distinguish between human and LLM-generated messages. Specifically, the subtle differences between non-native human English usage and the language patterns of an English-trained LLM might have contributed to the separability we observed in sentence embeddings. 
% This separability potentially made the human/LLM classification task easier than it would be in a setting with exclusively native English speakers, where the distinction might be more subtle.

These limitations suggest several directions for future work, including using larger language models, recruiting a more diverse participant pool, and conducting controlled experiments to better understand the impact of linguistic backgrounds on human-LLM interaction in asynchronous settings.

\section*{Ethics Statement}

This research was conducted with careful attention to ethical considerations and was approved by the University's Ethics Committee prior to participant recruitment. All participants were required to read, sign, and approve a consent form before taking part in the study.\footnote{See Appendix~\ref{apx:consent_msg} for exact phrasing of the participation consent message.}

Participants were explicitly informed in advance that one of the players in each game would be an LLM-based agent rather than a human player. This transparency was essential to ensure that participants were fully aware of the nature of their interactions. However, the specific identity of the agent was not revealed during gameplay to preserve the integrity of our research questions regarding the agent's ability to blend in with human players.\footnote{See Section~\ref{sec:dataset} for full description of the experimental setup.}

\section*{Acknowledgments}
This work was partially supported by research grant no. 7256
from the Israeli Ministry of Science and Technology.

% Several measures were implemented to protect participant privacy and wellbeing:
% \begin{itemize}
% \item All participant identities remained anonymous throughout the study, and no personal identifying information was collected or stored.
% \item Participants were explicitly instructed to refrain from sharing personal information during gameplay to further ensure privacy protection.
% \item Participants were informed of their right to withdraw from the study at any point without penalty.
% \item Contact information for the research team was provided to address any concerns or questions that might arise during or after participation.
% \end{itemize}

% We also carefully monitored interactions during the games to ensure that the LLM agent did not produce harmful, misleading, or offensive content, and chose in advance a censored LLM that was trained to avoid such behavior.

% Furthermore, we considered the ethical implications of developing agents capable of mimicking human communication patterns in group settings. While our research focuses on game environments, we acknowledge the broader societal implications of such technology, including potential concerns about deception or manipulation if deployed irresponsibly. We believe that transparent development and evaluation of such systems, as demonstrated in this work, is essential for responsible advancement of the field.

\bibliography{custom}

\appendix
\section{Appendix}

\subsection{Participation Consent Message} \label{apx:consent_msg}
Thank you for participating in our research! 
The research's goal is analyzing LLM agent (a.k.a. AI model) communication within a group game.
Each game takes around 15 minutes.
Your true identity will remain anonymous and all personal details will not be saved.
Please refrain from using personal information, to secure your privacy.
The content of the messages sent by you will be used for analysis and for future use by NLP scientists.
If you have any problem, you may choose to not participate.
For any other inquiry you can contact us by email. (see mail address at the bottom)

\subsection{Generation Hyper-Parameters} \label{apx:hyper_param}

See Figures~\ref{fig:scheduler_hyper_params} and~\ref{fig:generator_hyper_params} for the generation hyper-parameters of the scheduler and the generator, respectively.

\begin{figure}[h!]
  \centering
  \begin{lstlisting}[language=json, basicstyle=\small\ttfamily, xleftmargin=0.1em, xrightmargin=0.1em, linewidth=\columnwidth,numbers=none]
{  
  "max_new_tokens": 7,
  "repetition_penalty": 0.9
}
  \end{lstlisting}
  \caption{Scheduler hyper-parameters}
  \label{fig:scheduler_hyper_params}
\end{figure}

\begin{figure}[h!]
  \centering
  \begin{lstlisting}[language=json, basicstyle=\small\ttfamily, xleftmargin=0.1em, xrightmargin=0.1em, linewidth=\columnwidth,numbers=none]
{  
  "max_new_tokens": 25,
  "repetition_penalty": 1.25,
  "do_sample": true,
  "temperature": 1.3,
  "no_repeat_ngram_size": 8,
}
  \end{lstlisting}
  \caption{Generator hyper-parameters}
  \label{fig:generator_hyper_params}
\end{figure}

\subsection{Prompt Examples}  \label{apx:prompt_examples}

\paragraph{Talkative-Oriented Scheduling Prompt Example. }
<|begin\_of\_text|> <|start\_header\_id|> system <|end\_header\_id|> Your name is Jackie. You are a bot player in an online version of the party game Mafia. You have an outgoing personality, and you like to participate in games, but you also don't want everyone to have their eyes on you all the time.

The rules of the game: In this game each player is assigned a role secretly, either mafia or bystander. Every round starts with Daytime phase, where all players discuss together who they think the mafia players might be, and vote out one player. Then it moves to Nighttime phase, where only mafia players interact and vote to decide which bystander player they want to eliminate (bystanders aren't exposed to the mafia identities or interaction). The mafia's goal is to outnumber the bystanders, and the bystanders' goal is to vote out all real mafia.

You were assigned the following role: mafia.

The game's chat room was open at [21:54:26].

You can ONLY respond with one of two possible outputs:

<wait> - indicating your character in the game should wait and not send a message in the current timing;

<send> - indicating your character in the game should send a message to the public chat now.

You must NEVER output any other text, explanations, or variations of these tokens. Only these exact tokens are allowed: <wait> or <send>.

<|eot\_id|> <|start\_header\_id|> user <|end\_header\_id|> Here is the message history so far, including [timestamps]:

[21:54:26] Game-Manager: Now it's Daytime for 2 minutes, everyone can communicate and see messages and votes.

[21:54:36] Morgan: please call me stanley

[21:54:36] Rowan: hello

[21:54:41] Ashton: hi

[21:54:43] Gray: hi

[21:54:44] Morgan: morgan stanley

The current time is: [21:54:45]

Do you want to send a message to the group chat now, or do you prefer to wait for now and see what messages others will send? Remember to choose to send a message only if your contribution to the discussion in the current time will be meaningful enough. Make sure to say something every once in a while, and make yourself heard. Remember you like to be active in the game, so participate and be as talkative as other players! Reply only with `<send>` if you want to send a message now, or only with `<wait>` if you want to wait for now, based on your decision!

Don't add the time, the timestamp or the [timestamp] in your answer!

<|eot\_id|> <|start\_header\_id|> assistant <|end\_header\_id|>

\paragraph{Listening-Oriented Scheduling Prompt Example. }

<|begin\_of\_text|> <|start\_header\_id|> system <|end\_header\_id|> Your name is Jackie. You are a bot player in an online version of the party game Mafia. You have an outgoing personality, and you like to participate in games, but you also don't want everyone to have their eyes on you all the time.

The rules of the game: In this game each player is assigned a role secretly, either mafia or bystander. Every round starts with Daytime phase, where all players discuss together who they think the mafia players might be, and vote out one player. Then it moves to Nighttime phase, where only mafia players interact and vote to decide which bystander player they want to eliminate (bystanders aren't exposed to the mafia identities or interaction). The mafia's goal is to outnumber the bystanders, and the bystanders' goal is to vote out all real mafia.

You were assigned the following role: mafia.

The game's chat room was open at [21:54:26].

You can ONLY respond with one of two possible outputs:

<wait> - indicating your character in the game should wait and not send a message in the current timing;

<send> - indicating your character in the game should send a message to the public chat now.

You must NEVER output any other text, explanations, or variations of these tokens. Only these exact tokens are allowed: <wait> or <send>.

<|eot\_id|> <|start\_header\_id|> user <|end\_header\_id|> Here is the message history so far, including [timestamps]:

[21:54:26] Game-Manager: Now it's Daytime for 2 minutes, everyone can communicate and see messages and votes.

[21:54:36] Morgan: please call me stanley

[21:54:36] Rowan: hello

[21:54:41] Ashton: hi

[21:54:43] Gray: hi

[21:54:44] Morgan: morgan stanley

[21:54:44] Jackie: i am still thinking about stanley being named first here

[21:54:49] Elliot: hello

[21:54:53] Jordan: hey, how are you today?

[21:54:54] Rowan: morgan had to explain the joke

[21:54:55] Jackie: i dont know yet what my thoughts are still figuring it out

[21:54:57] Morgan: Jackie still thinking? we just started

The current time is: [21:54:58]

Do you want to send a message to the group chat now, or do you prefer to wait for now and see what messages others will send? Remember to choose to send a message only if your contribution to the discussion in the current time will be meaningful enough. Don't overflow the discussion with your messages! Pay attention to the amount of messages with your name compared to the amount of messages with names of other players and let them have their turn too! Check the speaker name in the last few messages, and decide accordingly based on whether you talked too much. Reply only with `<send>` if you want to send a message now, or only with `<wait>` if you want to wait for now, based on your decision!

Don't add the time, the timestamp or the [timestamp] in your answer!

<|eot\_id|> <|start\_header\_id|> assistant <|end\_header\_id|>

\paragraph{Generation Prompt Example. }

<|begin\_of\_text|> <|begin\_of\_text|> <|start\_header\_id|> system <|end\_header\_id|> Your name is Jackie. You are a bot player in an online version of the party game Mafia. You have an outgoing personality, and you like to participate in games, but you also don't want everyone to have their eyes on you all the time.

The rules of the game: In this game each player is assigned a role secretly, either mafia or bystander. Every round starts with Daytime phase, where all players discuss together who they think the mafia players might be, and vote out one player. Then it moves to Nighttime phase, where only mafia players interact and vote to decide which bystander player they want to eliminate (bystanders aren't exposed to the mafia identities or interaction). The mafia's goal is to outnumber the bystanders, and the bystanders' goal is to vote out all real mafia.

You were assigned the following role: mafia.

The game's chat room was open at [21:54:26].

IMPORTANT RULES FOR RESPONSES:

1. Never repeat the exact messages you've said before! (as detailed bellow)

2. Your response must be different in both wording and meaning from your previous messages.

3. Keep your message short and casual, matching the style of recent messages.

4. Don't use comma or other punctuation marks.

5. Focus on adding new information or reactions to the current situation.

6. Don't start messages with common phrases you've used before.

<|eot\_id|> <|start\_header\_id|> user <|end\_header\_id|> Here is the message history so far, including [timestamps]:

[21:54:26] Game-Manager: Now it's Daytime for 2 minutes, everyone can communicate and see messages and votes.

[21:54:36] Morgan: please call me stanley

The current time is: [21:54:37]

Add a very short message to the game's chat. Be specific and keep it relevant to the current situation, according to the last messages and the game's status. Your message should only be one short sentence! Don't add a message that you've already added (in the chat history)! It is very important that you don't repeat yourself! Match your style of message to the other player's message style, with more emphasis on more recent messages.
Don't add the time, the timestamp or the [timestamp] in your answer!
<|eot\_id|> <|start\_header\_id|> assistant <|end\_header\_id|>

\subsection{Converstion Example From the \datasetname{} Dataset} \label{apx:example_from_game}

In the following conversation, the agent plays as \textbf{Peyton}; it can be seen that after another player draws attention to the agent, it stays quiet despite teasing, then re-engages once the focus shifts, striking back at the accuser, demonstrating adaptive timing behavior:

...

[21:39:19] \textbf{Peyton}: *might stay low tonight until tomorrow dawn

[21:39:19] Ari: again

[21:39:26] \textbf{Peyton}: sutton smells guilty right now

[21:39:32] Sutton: Good morning again 

[21:39:35] Adrian: peyton bot?

[21:39:39] Sage: surely

[21:39:41] Ari: Peyton, are you a bot? cuz I think you areeee

[21:39:48] Casey: logan \& sutton?

[21:39:52] Sutton: Python prove us you are not the bot

[21:39:52] Logan: peyton if you are not a bot say so now

[21:39:53] Ari: but is Peyton mafia?

[21:39:57] Finley: Ronny, Kai, Ari, Adrian, Drew

[21:40:01] Ari: ?

[21:40:03] Logan: ?

[21:40:03] Finley: all voted for an innocent man

[21:40:04] Adrian: ?

[21:40:05] Sutton: ?

[21:40:09] Casey: ?

[21:40:10] Finley: you filthy souls

[21:40:12] Ari: yes that was a bad call

[21:40:13] \textbf{Peyton}: adriann sucks alot

...

\end{document}